\documentclass{aip-cp}

\usepackage[numbers]{natbib}
\usepackage{rotating}
\usepackage{graphicx}
\usepackage{bm}
\usepackage{amssymb}
\newcommand{\bra}[1]{\langle \, #1 \, |}
\newcommand{\ket}[1]{| \, #1 \, \rangle}
\newcommand{\kket}[1]{\, #1 \, \rangle}

\newcommand{\physdim}[1]{\hspace{1ex} (\mathrm{#1})}
\begin{document}

\title{Compositeness of Near Threshold Quasi-Bound States }

\author[aff1]{Yuki Kamiya\corref{cor1}}
\author[aff1]{Tetsuo Hyodo}

\affil[aff1]{ Yukawa Institute for Theoretical Physics, Kyoto University, Kyoto, 606-8502, Japan
}
\corresp[cor1]{Corresponding author: yuki.kamiya@yukawa.kyoto-u.ac.jp}

\maketitle

\begin{abstract}
We study the compositeness of unstable hadrons which lie near the two-hadron threshold. In the framework of the effective field theory, we derive the relation between the compositeness of stable bound states with observables. We then extend this relation for the quasi-bound states with finite decay width. A prescription to interpret the complex value of the compositeness is presented. With this method, we show that $\Lambda(1405)$ is dominated by the $\bar{K}N$ composite component and $a_0(980)$ have only small $\bar{K}K$ fraction.

\end{abstract}

\section{INTRODUCTION}

The discoveries of many $XYZ$ states in the heavy sector have made the study of hadrons enter a new stage~\cite{Swanson:2006st,Brambilla:2010cs}. Because these hadrons are found near the threshold of two mesons, contribution from the molecular state of two mesons is considered to be important for the internal structure of the $XYZ$ states. On the other hand, there are also other interpretations, such as the conventional quarkonia, tetra-quark states, gluon hybrids, and so on. It is therefore an urgent issue to reveal the internal structure of hadrons.

In general, the identification of the internal structure of hadrons is not straightforward, because the states with the same quantum numbers can mix with each other. A common approach is to construct a model with specific model space for the comparison with the experimental data. However, the contribution from the degrees of freedom which are not considered in the model can in principle be included in the model parameters. The connection between the model space and the described hadron is therefore indistinct. It is desirable to study the internal structure of hadrons directly from experimental observables without using specific models. 

It is shown in Ref.~\cite{Weinberg:1965zz} that the compositeness of a weakly binding state is determined only from the experimental observables in a model independent manner. Defining the compositeness $X$ as a probability to find the composite state in the physical bound state, we obtain the following weak-binding relation between the scattering length $a_0$ and the binding energy $B$;
\begin{equation}
	a_{0} = R\left\{ \frac{2X}{1+X} + {\mathcal O}\left(\frac{R_{\mathrm{typ}}}{R}\right)\right\}. \label{eq:comp-rel-bound}
\end{equation}
Here $\mu$ is the reduced mass and $R =1/\sqrt{2\mu B}$ is the length scale related to the binding energy $B$. $R_{\mathrm{typ}}$ is the typical length scale of the interaction of this system. This relation tells us that the scattering length $a_0$ is determined by the binding energy $B$ and the compositeness $X$ when the binding energy is small enough to satisfy $R_{\mathrm{typ}}/R\ll 1$. In other words, the compositeness of the weakly binding state can be calculated from the observables without using models. This relation is used to show that the deuteron is a proton-neutron composite system.

Because this relation is valid only for the stable states, it cannot be applied to unstable hadrons with decay modes. There are several attempts to study the compositeness of the unstable states~\cite{Baru:2003qq,Hyodo:2011qc,Aceti:2012dd,Hyodo:2013iga,Hyodo:2013nka,Sekihara:2014kya,Guo:2015daa}. Here we present the generalization of the weak binding relation~(\ref{eq:comp-rel-bound}) using the effective field theory (EFT)~\cite{Kamiya:2015aea}.

\section{MODEL-INDEPENDENT RELATIONS FOR BOUND STATES}

We first consider the single-channel $s$-wave scattering with a shallow bound state. Our interest is focused on the low energy physics near the threshold. We analyze this system with the effective non-relativistic quantum field theory with contact interactions~\cite{Kaplan:1996nv,Braaten:2007nq}
\begin{eqnarray}
H_{\mathrm{free}} &=&\int d\bm{r}
\biggl[\frac{1}{2 M} \mathbf{\nabla} \psi^\dagger \cdot\mathbf{\nabla} \psi +\frac{1}{2 m} \mathbf{\nabla} \phi^\dagger \cdot\mathbf{\nabla} \phi 
 + \frac{1}{2 M_{0}} \mathbf{\nabla}  B_0^\dagger \cdot{\mathbf \nabla} B_0 +  \nu_0 B_0^\dagger B_0
\biggr] ,\\
H_{\mathrm{int}} &=& \int d\bm{r}
\left[
g_{0} \left( B_0^\dagger \phi\psi + \psi^\dagger\phi^\dagger B_0 \right) + \lambda_0 \psi^\dagger\phi^\dagger \phi\psi
\right].
\end{eqnarray}
Here we take $\hbar =1$. We consider that this EFT is applicable below the cutoff momentum scale $\Lambda$. The cutoff scale is related to the typical length scale of the interaction $R_{\mathrm{typ}}$ as $ \Lambda\sim 1/R_{\mathrm{typ}}$. Now we consider that the full Hamiltonian has a discrete eigenstate $|B\rangle$ with the binding energy $B$ in the same quantum numbers with the two-body $\psi\phi$ system as
\begin{eqnarray}
   H\ket{B}=-B\ket{B}
\end{eqnarray}
From the phase symmetry of the Hamiltonian, it can be shown that the completeness relation in this sector is spanned by the eigenstates of $H_{\mathrm{free}}$: the scattering states $\ket{\bm{p}}=\tilde{\psi}^{\dag}(\bm{p})\tilde{\phi}^{\dag}(-\bm{p})/\sqrt{\mathcal{V}}\ket{0}$ and the discrete state $\ket{B_{0}}=\tilde{B}_{0}^{\dag}(\bm{0})/\sqrt{\mathcal{V}}\ket{0}$ with the creation operators $\tilde{\psi}^{\dag}(\bm{p})$, the vacuum $\ket{0}$, and $\mathcal{V}=(2\pi)^{3}\delta^{3}(\bm{0})$. This enables us to write the bound state $|B\rangle$ as a linear combination of these states,
\begin{eqnarray}
   \ket{B} 
   =c\ket{B_{0}}+\int\frac{d\bm{p}}{(2\pi)^{3}} \chi(\bm{p})\ket{\bm{p}}
   \label{eq:twobody}.
\end{eqnarray}
Now we define the compositeness $X$ (the elementariness $Z$) as the probability to find the scattering (discrete) state in the bound state;
\begin{eqnarray}
Z \equiv |\bra{B_{0}}\kket{B}|^{2}=|c|^2,\quad 
X \equiv \int \frac{d\bm{p}}{(2\pi)^{3}} |\bra{\bm{p}}\kket{B}|^{2}= \int \frac{d\bm{p}}{(2\pi)^3} |\chi(\bm{p})|^2 .
\label{eq:normalization}
\end{eqnarray}
With the normalization condition of the bound state $\langle B|B\rangle = 1$, $Z$ and $X$ are shown to satisfy the following relations:
\begin{eqnarray}
Z+X =1,\quad Z,X \in [0,1] .
\label{eq:ZXbound}
\end{eqnarray}
This ensures the probabilistic interpretation of $X$ and $Z$.

The exact forward scattering amplitude of the $\psi \phi$ system $f(E)$ is obtained as 
\begin{eqnarray}
	f(E) =-\frac{\mu}{2\pi} \frac{1}{\left[v(E)\right]^{-1} - G(E) } \label{eq:amp-bound},
\end{eqnarray}
with $\mu\equiv Mm/(M+m)$ and 
\begin{eqnarray}
v(E) =\lambda_0 + \frac{g_0^2}{E - \nu_0}, \quad
G(E) = \frac{1}{2\pi^2} 
\int_{0}^{\Lambda}dp \frac{p^{2}}{E-p^2/(2\mu)+i0^{+}}.
\end{eqnarray}
The bound state condition is obtained from Eq. (\ref{eq:amp-bound}) as
\begin{eqnarray}
G(-B)v(-B) = 1 \label{eq:pole-cond}.
\end{eqnarray}
The compositeness $X$ and the elementariness $Z$ can be expressed by $v$ and $G$ as~\cite{Sekihara:2014kya}
\begin{eqnarray}
X  = \frac{1}{1+ G^2(-B) v^\prime (-B)\left[ G^\prime (-B)\right]^{-1}} , \quad 
Z  = \frac{1}{1+ v^2(-B) \left[v^\prime (-B)\right]^{-1}G^\prime (-B)} \label{eq:X-bound} ,
\end{eqnarray}
where $A^\prime(E)$ denotes the derivative of $A(E)$ with respect to the energy $E$.

We then expand the scattering length $a_0 \equiv -f(0)$ in powers of $1/R$. Using Eq.~(\ref{eq:X-bound}) and the bound state condition~(\ref{eq:pole-cond}), we find that the coefficient of the leading order term $\mathcal{O}(R)$ is expressed only by the compositeness $X$. This verifies the weak-binding relation~(\ref{eq:comp-rel-bound}). The higher order terms are suppressed by $R_{\mathrm{typ}}/R$ compared with the leading order term and depends explicitly on the cutoff $\Lambda$. The details of the short range behavior of the interaction is reflected in the higher order terms. Thus, when the binding energy is sufficiently small and the higher order terms can be neglected, the compositeness $X$ is determined only from the observable quantities, the scattering length $a_{0}$ and the binding energy $B$. We have to notice that the compositeness $X$ is a model-dependent quantity, when the binding energy is not small. It is shown that the compositeness $X$ in expression~(\ref{eq:X-bound}) is not invariant under the change of the cutoff scale $\Lambda \to \Lambda + \delta \Lambda$~\cite{Hyodo:2015weg}. The weak-binding limit is the exceptional case where the model dependence of $X$ is suppressed in the expansion of Eq.~(\ref{eq:comp-rel-bound}), and the structure of the bound state can be determined from the observables.

\section{MODEL-INDEPENDENT RELATIONS FOR QUASI-BOUND STATES} \label{sec:quasi}

To generalize the weak-binding relation to the unstable quasi-bound states, we add an additional two-body channel to the previous system. The corresponding EFT can be described by the Hamiltonian $H = H_{\mathrm{free}} + H_{\mathrm{int}}$;
\begin{eqnarray}
H_{\mathrm
	{free}} &=&\int d\bm{r} \Biggl[\sum_{i=1,2}\frac{1}{2 M_i} \mathbf{\nabla} \psi_i^\dagger \cdot\mathbf{\nabla} \psi_i +\sum_{i=1,2}\frac{1}{2 m_i} \mathbf{\nabla} \phi_i^\dagger \cdot\mathbf{\nabla} \phi_i + \frac{1}{2M_{0}} \mathbf{\nabla}  B_0^\dagger \cdot{\mathbf \nabla} B_0 -\nu_\psi \psi_2^\dagger \psi_2-\nu_\phi \phi_2^\dagger \phi_2+\nu_0 B_0^\dagger B_0 \Biggr],\\
H_{\mathrm{int}} &=&\int d\bm{r} \left[\sum_{i=1,2} g_{0,i} \left( B_0^\dagger \phi_i\psi_i + \psi_i^\dagger\phi_i^\dagger B_0 \right) + \sum_{i,j=1,2} \lambda_{0,ij} \psi_j^\dagger\phi_j^\dagger \phi_i\psi_i\right],
\end{eqnarray}
with $\lambda_{0,12} = \lambda_{0,21}$. Setting the energy difference of the threshold as $\nu \equiv  \nu_\psi + \nu_\phi > 0$, we find that the threshold of the channel 2 ($\psi_2 \phi_2$) is lower than that of channel 1 ($\psi_1 \phi_1$). We consider the full Hamiltonian has a quasi-bound state $\ket{QB}$ with the complex eigenenergy $E_{QB}\in \mathbb{C}$ as 
\begin{eqnarray}
  H\ket{QB}=E_{QB}\ket{QB} ,
\end{eqnarray}
which can be expanded as a linear combination
\begin{eqnarray}
  \ket{QB} 
  =c\ket{B_{0}}+\sum_{i=1,2}\int\frac{d\bm{p}}{(2\pi)^{3}} 
  \chi_i(\bm{p})\ket{\bm{p}_i}
  \label{eq:twobody2} ,
\end{eqnarray}
with $\ket{\bm{p}_i}=\tilde{\psi}_i^{\dagger}(\bm{p})\tilde{\phi}_i^{\dagger}(-\bm{p})/\sqrt{\mathcal{V}}\ket{0}$.

The wave function of an unstable state diverges at large distance and cannot be normalized. To normalize the unstable state, we introduce the Gamow state $|\overline{QB} \rangle \equiv |QB \rangle ^*$. With the analytic continuation using the convergence factor, one can show that the normalization condition $\langle \overline{QB}|QB\rangle = 1$ is well defined. The definition of $X$ and $Z$ is then given by
\begin{equation}
Z\equiv  \bra{\overline{QB}}\kket{B_0}\bra{B_{0}}\kket{QB}= c^2,
 \quad 
 X_i = \int \frac{d\bm{p}}{(2\pi)^{3}} \bra{\overline{QB}}\kket{\bm{p}_i}\bra{\bm{p}_i}\kket{QB}=\int \frac{d\bm{p}}{(2\pi)^{3}} \chi_i^2(\bm{p}),\label{eq:X-def-quasi} 
\end{equation}
which satisfy
\begin{equation}
Z+X_1+X_2=1,\quad Z,X_i \in \mathbb{C} .\label{eq:sum-rule-quasi} 
\end{equation}
The complex nature of $X_{i}$ and $Z$ reflects the fact that the expectation value of an arbitrary operator for the unstable state becomes complex. Nevertheless, the sum rule~(\ref{eq:sum-rule-quasi}) is guaranteed owing to the normalization of $\ket{QB}$.

The forward scattering amplitude of the channel 1 is given by
\begin{eqnarray}
	f_{11}(E) =-\frac{\mu_{1}}{2\pi} \frac{1}{\left[v(E)\right]^{-1} - G_{1}(E) }
\end{eqnarray}
with
\begin{equation}
v(E) = \lambda_{0,11}+\frac{g_{0,1}^{2}}{E-\nu_{0}}
   +\frac{\left(\lambda_{0,12}+\frac{g_{0,1}g_{0,2}}{E-\nu_{0}}\right)^{2}}
   {[G_2(E)]^{-1}-(\lambda_{0,22}+\frac{g_{0,2}^{2}}{E-\nu_{0}})} 
   \label{eq:vE},\quad 
G_i(E)
   = \frac{1}{2\pi^2} 
\int_{0}^{\Lambda}dp \frac{p^{2}}{E-p^2/(2\mu_i)+\delta_{i,2}\nu+i0^{+}}
\end{equation}
with $\mu_i = m_iM_i/(m_i+M_i)$. Using these expression, the compositeness of the channel 1, $X_1$, can be written as
\begin{eqnarray}
X_1 = \frac{1}{1 + G_{1}^{2}(E_{QB})v^\prime(E_{QB}) [G_1^\prime(E_{QB})]^{-1}}.
\end{eqnarray}

Now we redefine $R$ by 
\begin{eqnarray}
R \equiv 1/\sqrt{-2\mu_1 E_{QB}}.
\end{eqnarray}
Expanding the scattering length $a_0$ in powers of $1/R$, we obtain the weak-binding relation for the quasi-bound state;
\begin{eqnarray}
a_0
=  R \Biggl\{\frac{2X_1}{1+X_1} + {\mathcal O}\left(\left|\frac{R_{\mathrm{typ}}}{R}\right| \right) + \sqrt{\frac{\mu_2^{3}}{\mu_1^{3}}} \mathcal{O} \left( \left| \frac{l}{R} \right|^{3}\right) \Biggr\}
\label{eq:comp-rel-quasi},
\end{eqnarray}
where $l=1/\sqrt{2\mu_{1}\nu}$. The first two terms arise in the same manner with the bound state case. The last term originates in the expansion of the $G_{2}(E)$, which contains the additional length scale $l$. When the energy difference between the thresholds $\nu$ is large, the length scale $l$ is small. To determine the compositeness $X$ model independently, in addition to the condition $|R_{\mathrm{typ}}/R|\ll 1$, the eigenenergy should satisfy $|l/R|^3 \ll 1$. The latter condition is met when the decay channel 2 is sufficiently separated from the channel 1 with respect to the magnitude of the eigenenergy. It is clear from the derivation that the relation~(\ref{eq:comp-rel-quasi}) holds also for the states with Re $E_{QB}>0$.

When $X_{1}$ is determined model independently, we can also determine $Z+X_{2}$ from Eq.~(\ref{eq:sum-rule-quasi}). In this case, however, separation of $Z$ and $X_{2}$ is model dependent. In the following, we rewrite $X_1$ as $X$ and $Z+X_2$ as $Z$. The sum rule is then given by $Z+X=1$ and $Z$ represents any contributions other than the channel 1.

\section{INTERPRETATION OF COMPLEX COMPOSITENESS}

From the definition~(\ref{eq:X-def-quasi}), the compositeness of the quasi-bound state is obtained as a complex number, which cannot be directly interpreted as a probability. There have been several works for the interpretation of the complex compositeness. For example, in Ref. \cite{Aceti:2014ala}, it is suggested to consider the real part of the compositeness as a probability, because the sum rule $\mathrm{Re}\phantom{x}Z+\mathrm{Re}\phantom{x}X=1$ holds. However, Re $X$ can be negative or larger than one.

Here we introduce two real quantities, $\tilde{X}$ and $\tilde{Z}$, which are to be regarded as probabilities to find the composite component and the other components in the physical quasi-bound state, respectively. For the meaningful probabilistic interpretation, the following conditions should be satisfied:
\begin{eqnarray}
	\tilde{Z}+\tilde{X} = 1 ,\quad \tilde{Z},\tilde{X} \in [0,1] . \label{eq:cond-tilde-X}
\end{eqnarray}
If the width of the quasi-bound state is small, we expect that its wave function becomes similar to the stable bound state~\cite{Sekihara:2014kya}. In the small width limit, therefore, the values of $\tilde{Z}$ and $\tilde{X}$ should approach $Z$ and $X$, respectively. On the other hand, if the imaginary part of $X$ or the magnitude of the real part is very large, there is no corresponding bound state wave function. In such cases, the compositeness $X$ should not be regarded as the measure of the internal structure. To quantify the clarity of the interpretation, we introduce another real quantity $U\geq 0$, such that $U$ increases when the interpretation is unclear. Based on these considerations, we propose to define $\tilde{X},\tilde{Z},U$ from $Z,X$ as
\begin{eqnarray}
    \tilde{Z} \equiv \frac{1 - |X| + |Z|}{2},
    \quad \tilde{X} \equiv \frac{1 - |Z| + |X|}{2}, \quad 
    U \equiv |Z| +|X| -1 \label{eq:kaisyaku2-U} .
\end{eqnarray}
where the definition of $U$ is inspired by Ref.~\cite{PL33B.547}. These quantities satisfy Eq.~(\ref{eq:cond-tilde-X}). When Eq.~(\ref{eq:ZXbound}) is satisfied, we obtain $\tilde{Z}=Z$, $\tilde{X}=X$ and $U=0$. When $U$ is large, the deviation from the bound state case becomes significant. We thus interpret $\tilde{X}$ and $\tilde{Z}$ as probabilities, only when the uncertainty $U$ is small.

\section{APPLICATIONS TO EXOTIC HADRONS}

Now we use Eq.~(\ref{eq:comp-rel-quasi}) to study the compositeness of hadronic states. For the application, it is required that (i) the scattering length and the eigenenergy are known to some extent, and
(ii) the quasi-bound state appears near the $s$-wave two-hadron threshold.

\begin{table}[bt]
			\begin{tabular}{llllcl}
				\hline
				Ref. & $E_{QB}\physdim{MeV}$ & $a_0 \physdim{fm} $ & $X_{\bar{K}N}$ & $\tilde{X}_{\bar{K}N}$ & $U$   \\  \hline
				\cite{Ikeda:2012au}  & $-10-i26$ & $1.39 - i 0.85$ 
				& $1.2+i0.1$ & $1.0$ & $0.5$   \\ 
				\cite{Mai:2012dt}  & $-\phantom{0}4-i\phantom{0}8$ & $1.81-i0.92$ 
				& $0.6+i0.1$ & $0.6$ & $0.0$  \\ 
				\cite{Guo:2012vv}  & $-13-i20$ & $1.30-i0.85$ 
				& $0.9-i0.2$ & $0.9$ & $0.1$   \\
				\cite{Mai:2014xna}  & $\phantom{-0}2-i10$ & $1.21-i1.47$ 
				& $0.6+i0.0$ & $0.6$ & $0.0$  \\ 
				\cite{Mai:2014xna}  & $-\phantom{0}3-i12$ & $1.52-i1.85$ 
				& $1.0+i0.5$ & $0.8$ & $0.6$  \\ \hline
			\end{tabular} 
			\caption{Properties and results for $\Lambda (1405)$. Shown are the eigenenergy $E_{QB}$, $\bar{K}N(I=0)$ scattering length $a_{0}$, the $\bar{K}N$ compositeness $X_{\bar{K}N}$ and $\tilde{X}_{\bar{K}N}$, uncertainty $U$. The scattering length is defined as $a_{0}=-f(E=0)$.}
			\label{tab:Lambda}
\end{table}

First we discuss the structure of $\Lambda(1405)$. This state lies close to the $\bar{K}N$ threshold and decays into the $\pi\Sigma$ channel. To apply the near-threshold formula, we focus on the higher energy one of the two poles associated with this resonance~\cite{Jido:2003cb}. Recently the eigenenergy and scattering length of this state are obtained by analyses based on chiral effective theory~\cite{Ikeda:2011pi,Ikeda:2012au,Mai:2012dt,Guo:2012vv,Mai:2014xna}. From the eigenenergies found in these studies, we find that the value of $R$ takes $|R| \gtrsim 1.5\hspace{1ex}\mathrm{fm}$. Then the correction terms are found to be small, $|R_{\mathrm{typ}}/R|\lesssim 0.17$ and $|l/R|^3 \lesssim 0.14$, where the typical length scale of $\bar{K}N$ interaction $R_{\mathrm{typ}}$ is estimated from $\rho$ meson exchange. We calculate $X_{\bar{K}N}$, $\tilde{X}_{\bar{K}N}$ and $U$ from the $\bar{K}N(I=0)$ scattering length and the eigenenergy. The results are summarized in Table~\ref{tab:Lambda}. In all cases, $U$ is not large and $\tilde{X}_{\bar{K}N}$ is obtained around unity. Thus we conclude that $\Lambda(1405)$ is dominated by the $\bar{K}N$ composite component.

\begin{table}[bt]
			\begin{tabular}{llllcl}
				\hline
				Ref. & $E_{QB} \physdim{MeV}$ & $a_0 \physdim{fm} $ & $X_{\bar{K}K}$ & $\tilde{X}_{\bar{K}K}$ & $U$   \\  \hline
				\cite{Adams:2011sq}  & $31-i70$ & $-0.03 - i 0.53$ 
				& $0.2-i0.2$ & $0.3$ & $0.1$   \\ 
				\cite{Ambrosino:2009py} & $\phantom{0}3-i25$ & $\phantom{-}0.17 - i 0.77$ 
				& $0.2-i0.2$ & $0.2$ & $0.1$   \\ 
				\cite{Bugg:2008ig} & $\phantom{0}9-i36$ & $\phantom{-}0.05 - i 0.63$ 
				& $0.2-i0.2$ & $0.2$ & $0.1$   \\ 
				\cite{Achasov:2000ku} & $14-i\phantom{0}5$ & $-0.13 - i 2.19$ 
				& $0.8-i0.4$ & $0.7$ & $0.3$   \\ 
				\cite{Teige:1996fi} & $15-i29$ & $-0.13 - i 0.52$ 
				& $0.1-i0.2$ & $0.1$ & $0.1$   \\ \hline
			\end{tabular} 
			\caption{Properties and results for $a_{0} (980)$. Shown are the eigenenergy $E_{QB}$, $\bar{K}K(I=1)$ scattering length $a_{0}$, the $\bar{K}K$ compositeness $X_{\bar{K}K}$ and $\tilde{X}_{\bar{K}K}$.}
			\label{tab:a0}
\end{table}

\begin{table}[bt]
			\begin{tabular}{llllcl}
				\hline
				Ref. & $E_{QB} \physdim{MeV}$ & $a_0 \physdim{fm} $ & $X_{\bar{K}K}$ & $\tilde{X}_{\bar{K}K}$ & $U$   \\  \hline
				\cite{Aaltonen:2011nk}  & $\phantom{-}19-i30$ & $0.02 - i 0.95$ 
				& $0.3-i0.3$ & $0.4$ & $0.2$   \\ 
				\cite{Ambrosino:2005wk}  & $-\phantom{0}6-i10$ & $0.84 - i 0.85$ 
				& $0.3-i0.1$ & $0.3$ & $0.0$   \\ 
				\cite{Garmash:2005rv}  & $-\phantom{0}8-i28$ & $0.64 - i 0.83$ 
				& $0.4-i0.2$ & $0.4$ & $0.1$  \\ 
				\cite{Ablikim:2004wn}  & $\phantom{-}10-i18$ & $0.51 - i 1.58$ 
				& $0.7-i0.3$ & $0.6$ & $0.1$ \\ 
				\cite{Link:2004wx}  & $-10-i29$ & $0.49 - i 0.67$ 
				& $0.3-i0.1$ & $0.3$ & $0.0$ \\ 
				\cite{Achasov:2000ym}  & $\phantom{-}10-i\phantom{0}7$ & $0.52 - i 2.41$ 
				& $0.9-i0.2$ & $0.9$ & $0.1$ \\ \hline
			\end{tabular} 
			\caption{Properties and results for $f_{0} (980)$. Shown are the eigenenergy $E_{QB}$, $\bar{K}K(I=0)$ scattering length $a_{0}$, the $\bar{K}K$ compositeness $X_{\bar{K}K}$ and $\tilde{X}_{\bar{K}K}$, uncertainty $U$.}
			\label{tab:f0}
\end{table}

Next we discuss the scalar mesons $a_0(980)$ and $f_0(980)$ which appear near the $\bar{K}K$ threshold. $a_0(980)$ and $f_0(980)$ decay into the $\pi\eta$ channel and the $\pi\pi$ channel, respectively. With analyses of the recent experiments, the Flatte parameters of the $\bar{K}K$ scattering amplitude are determined~\cite{Adams:2011sq,Ambrosino:2009py,Bugg:2008ig,Achasov:2000ku,Teige:1996fi,Aaltonen:2011nk,Ambrosino:2005wk,Garmash:2005rv,Ablikim:2004wn,Link:2004wx,Achasov:2000ym}. The eigenenergy and threshold parameters are calculated from the Flatte parameters. Except for Ref.~\cite{Achasov:2000ku}, the eigenenergy  satisfies $|R_{\mathrm{typ}}/R|\lesssim 0.17$ and $|l/R|^3\lesssim 0.04$, where $R_{\mathrm{typ}}$ is estimated from the $\rho$ meson exchange again. From Ref.~\cite{Achasov:2000ku}, $|R_{\mathrm{typ}}/R|\sim 0.25$, $|l/R|^3\sim 0.13$ are obtained. The calculated results of compositeness for $a_0(980)$ [$f_0(980)$] are listed in Table~\ref{tab:a0} (\ref{tab:f0}). For $a_0(980)$, we see that $U$ is small enough for all cases. The result of $\tilde{X}_{\bar{K}K}$ is close to zero, except for Ref.~\cite{Achasov:2000ku}. Considering the large experimental uncertainty in the Flatte parameters of Ref.~\cite{Achasov:2000ku}, we conclude that $a_0(980)$ has only small $\bar{K}K$ composite component. For $f_0(980)$, though $U$ is small for all cases, the value of $\tilde{X}_{\bar{K}K}$ is scattered, because of the experimental uncertainty. To make a conclusive interpretation of the nature of $f_{0}(980)$, we need more accurate input data.

The $\bar{K}N$ compositeness of $\Lambda(1405)$ and the $\bar{K}K$ compositeness of the scalar mesons are studied in Ref.~\cite{Sekihara:2014kya,Sekihara:2014qxa} based on the chiral unitary model. The evaluated values of the compositeness in these studies are in good agreement with the model-independent determination. $\bar{K}N$ composite dominance of the $\Lambda(1405)$ has been indicated by the lattice calculation~\cite{Hall:2014uca} and by the analysis with the realistic $\bar{K}N$ potential~\cite{Miyahara:2015bya}. The conclusion of these works are strongly supported by the present model-independent analysis.

\section{SUMMARY}

We have presented the derivation of the weak-binding relation of the bound state based on the non-relativistic effective field theory. The weak-binding relation is then generalized to the quasi-bound states. The generalized relation clarifies the condition to neglect the contribution of the decay channel. When the magnitude of the eigenenergy of the quasi-bound state is small enough to neglect the model-dependent correction terms, it is possible to determine the compositeness $X$ of the quasi-bound state from the scattering length and the eigenenergy. We have suggested the prescription to interpret the complex compositeness $X$ by defining new real quantities $\tilde{X}$ and $U$. These quantities allow us to discuss the internal structure of the quasi-bound states with the probabilistic interpretation. As applications of this method, we have discussed the internal structure of $\Lambda(1405)$ and the scalar mesons. 
It is concluded that $\Lambda(1405)$ is dominated by the $\bar{K}N$ composite component and $a_0(980)$ have only small $\bar{K}K$ fraction.

\section{ACKNOWLEDGMENTS}
This work is supported in part by JSPS KAKENHI Grants No. 24740152 and by the Yukawa International Program for Quark-Hadron Sciences (YIPQS).



\begin{thebibliography}{30}
	

%
%
%
%
%
%
	
	\bibitem{Swanson:2006st}
	E.~S. Swanson,
	\newblock Phys. Rept. {\bf 429}, 243 (2006).
	
	\bibitem{Brambilla:2010cs}
	N.~Brambilla {\em et~al.},
	\newblock Eur. Phys. J. C {\bf 71}, 1534 (2011).

	
%
	
%
%

	

	
	\bibitem{Weinberg:1965zz}
	S.~Weinberg,
	\newblock Phys. Rev. {\bf 137}, B672 (1965).
	

	
	\bibitem{Baru:2003qq}
	V.~Baru, J.~Haidenbauer, C.~Hanhart, Y.~Kalashnikova and A.~E. Kudryavtsev,
	\newblock Phys. Lett. B {\bf 586}, 53 (2004).
	
	\bibitem{Hyodo:2011qc}
	T.~Hyodo, D.~Jido and A.~Hosaka,
	\newblock Phys. Rev. C {\bf 85}, 015201 (2012).
	
	\bibitem{Aceti:2012dd}
	F.~Aceti and E.~Oset,
	\newblock Phys. Rev. D {\bf 86}, 014012 (2012).
	
	\bibitem{Hyodo:2013iga}
	T.~Hyodo,
	\newblock Phys. Rev. Lett. {\bf 111}, 132002 (2013).
	
	\bibitem{Hyodo:2013nka}
	T.~Hyodo,
	\newblock Int. J. Mod. Phys. A {\bf 28}, 1330045 (2013).

	\bibitem{Sekihara:2014kya}
	T.~Sekihara, T.~Hyodo and D.~Jido,
	\newblock PTEP {\bf 2015}, 063D04 (2014).
	
	\bibitem{Guo:2015daa} 
	Z.~H.~Guo and J.~A.~Oller,
	arXiv:1508.06400 [hep-ph].
	
\bibitem{Kamiya:2015aea}
Y.~Kamiya and T.~Hyodo,
 arXiv:1509.00146 [hep-ph].

	
%
	
%
%
%
%
%
%
%
	\bibitem{Kaplan:1996nv}
	D.~B. Kaplan,
	\newblock Nucl. Phys. B {\bf 494}, 471 (1997).

	\bibitem{Braaten:2007nq}
	E.~Braaten, M.~Kusunoki and D.~Zhang,
	\newblock Annals Phys. {\bf 323}, 1770 (2008).
%
	\bibitem{Hyodo:2015weg} 
	T.~Hyodo,
	arXiv:1511.00870 [hep-ph].
%
%
	
	\bibitem{Aceti:2014ala}
	F.~Aceti, L.~Dai, L.~Geng, E.~Oset and Y.~Zhang,
	\newblock Eur. Phys. J. A {\bf 50}, 57 (2014).

	\bibitem{PL33B.547}
	T.~Berggren,
	\newblock Phys. Lett. {\bf 33B}, 547 (1970).
	
%

	\bibitem{Jido:2003cb} 
	D.~Jido, J.~A.~Oller, E.~Oset, A.~Ramos and U.~G.~Meissner,
	Nucl.\ Phys.\ A {\bf 725}, 181 (2003)
	
	\bibitem{Ikeda:2011pi}
	Y.~Ikeda, T.~Hyodo and W.~Weise,
	\newblock Phys. Lett. B {\bf 706}, 63 (2011).
	
	\bibitem{Ikeda:2012au}
	Y.~Ikeda, T.~Hyodo and W.~Weise,
	\newblock Nucl. Phys. A {\bf 881}, 98 (2012).
	
	\bibitem{Mai:2012dt}
	M.~Mai and U.-G. Meissner,
	\newblock Nucl. Phys. A {\bf 900}, 51  (2013).
	
	\bibitem{Guo:2012vv}
	Z.-H. Guo and J.A.~Oller,
	\newblock Phys. Rev. C {\bf 87}, 035202 (2013).
	
	\bibitem{Mai:2014xna}
	M.~Mai and U.-G. Mei{\ss}ner,
	\newblock Eur. Phys. J. A {\bf 51}, 30 (2015).
		
	\bibitem{Adams:2011sq}
	CLEO, G.~S. Adams {\em et~al.},
	\newblock Phys. Rev. D {\bf 84}, 112009 (2011).
	
	\bibitem{Ambrosino:2009py}
	KLOE, F.~Ambrosino {\em et~al.},
	\newblock Phys. Lett. B {\bf 681}, 5 (2009).
	
	\bibitem{Bugg:2008ig}
	D.~V. Bugg,
	\newblock Phys. Rev. D {\bf 78}, 074023 (2008).
	
	\bibitem{Achasov:2000ku}
	M.~N. Achasov {\em et~al.},
	\newblock Phys. Lett. B {\bf 479}, 53 (2000).
	
	\bibitem{Teige:1996fi}
	E852, S.~Teige {\em et~al.},
	\newblock Phys. Rev. D {\bf 59}, 012001 (1999).
	
	\bibitem{Aaltonen:2011nk}
	CDF, T.~Aaltonen {\em et~al.},
	\newblock Phys. Rev. D {\bf 84}, 052012 (2011).
	
	\bibitem{Ambrosino:2005wk}
	KLOE, F.~Ambrosino {\em et~al.},
	\newblock Phys. Lett. B {\bf 634}, 148 (2006).
	
	\bibitem{Garmash:2005rv}
	Belle, A.~Garmash {\em et~al.},
	\newblock Phys. Rev. Lett. {\bf 96}, 251803 (2006).
	
	\bibitem{Ablikim:2004wn}
	BES, M.~Ablikim {\em et~al.},
	\newblock Phys. Lett. B {\bf 607}, 243 (2005).
	
	\bibitem{Link:2004wx}
	FOCUS, J.~M. Link {\em et~al.},
	\newblock Phys. Lett. B {\bf 610}, 225 (2005).
	
	\bibitem{Achasov:2000ym}
	M.~N. Achasov {\em et~al.},
	\newblock Phys. Lett. B {\bf 485}, 349 (2000).

	\bibitem{Sekihara:2014qxa}
	T.~Sekihara and S.~Kumano,
	\newblock Phys. Rev. D {\bf 92}, 034010 (2015).

	\bibitem{Hall:2014uca}
	J.~M.~M. Hall {\em et~al.},
	\newblock Phys. Rev. Lett. {\bf 114}, 132002 (2015).

	\bibitem{Miyahara:2015bya}
	K.~Miyahara and T.~Hyodo,
	\newblock arXiv:1506.05724 [nucl-th].
	
%
\end{thebibliography}

\end{document}